\documentclass[twocolumn,twoside,slac]{revtex4}
\usepackage{graphicx}
\usepackage{fancyhdr}
\begin{document}

\title{The BTeV Software Tutorial Suite}

%

\author{R. K. Kutschke}
\affiliation{Fermi National Accelerator Laboratory, Batavia, IL 60510, USA}

\begin{abstract}
The BTeV Collaboration is starting to develop its 
C++ based offline software suite, an integral part of which
is a series of tutorials.  These tutorials are targeted
at a diverse audience, including new graduate students, experienced
physicists with little or no C++ experience, those with just enough
C++ to be dangerous, and experts who need only an overview of the 
available tools.  The tutorials must both teach C++ in general and the BTeV 
specific tools in particular. Finally, they must teach physicists 
how to find and use the detailed documentation. This report
will review the status of the BTeV experiment, give an overview of
the plans for and the state of the software and will then describe
the plans for the tutorial suite.
\end{abstract}

\maketitle

\thispagestyle{fancy}


\section{INTRODUCTION}

 The BTeV experiment is designed to make comprehensive and precise 
studies of  CP violation, flavor mixing and rare decays in
the fields of beauty and charm physics, both important components
in the study of flavor physics.  This broader field
includes the integration of beauty and charm physics with
kaon and neutrino physics and with some cosmological physics.   
A related physics goal is to perform an extensive search for
physics beyond the standard model, both by searches for rare or
forbidden processes and by precision self consistency tests of
a large body of measurements.  The proposed
detector is a forward spectrometer which will be built at 
the CZero interaction region of the Tevatron $p\bar{p}$ collider 
at Fermilab.  The Fermilab directorate has given stage I approval
to BTeV and final decisions on the funding profile and the
construction timetable are anticipated soon.  Current planning 
is for construction to begin in late 2004, with physics running
to begin in late 2008.   The collaboration is planning a staged
installation of the detector over a period of several years and
one can imagine that the first engineering runs on a partly
installed subsystem might take place as early as 2006.

A discussion of BTeV's physics reach and details about
the design of the spectrometer may be found on the
BTeV web site~\cite{BTeV:web}, in particular on the page
which links to the Proposal, the preliminary 
Technical Design Report (TDR) and related
documents~\cite{BTeV:docs}.  An excellent review of heavy
flavor physics at hadron colliders~\cite{BRunII} is
also available.

  The sorts of physics analyses to be performed will be similar
to the beauty and charm physics studies performed at the $e^+e^-$
B-factories and to the charm physics studies performed at the
last generation of fixed target detectors.  
That is, the two main types of analyses will be full reconstruction of 
exclusive final states and partial reconstruction of exclusive
final states using the line of flight as a constraint.  Unlike
the $e^+e^-$ B-factories, however, BTeV's
data rates and data volumes will challenge the state of the art
in data acquisition (DAQ), triggering and computing.  In this
way BTeV will be more like the current and next generations of
hadron colliders.

\section{OVERVIEW OF THE BTEV SOFTWARE}

 This section will discuss some unique aspects of the
BTeV software, overview its history and discuss the
present status and future plans. This will set
the stage for a discussion of the tutorial suite.

\subsection{Offline vs Online}

Over the past few decades an important trend in
High Energy Physics (HEP) has been that software has
moved to ever lower levels of the trigger system.
Moreover the previous distinctions between online 
trigger software and offline reconstruction
software are increasingly blurred.

BTeV will take the next step in this evolution by
performing, for every beam crossing,
track and vertex reconstruction 
at the lowest level of the trigger system, Level~1.
The Level~1 trigger decision will be based on
evidence for tracks which are detached from a
primary interaction vertex.  The Level~1 trigger algorithm 
uses only hits from the pixel vertex detector and it
must perform robustly even when each beam crossing
contains several background interactions.
One key to making this work is the extremely low
occupancy of the pixel detector system, which reduces
the combinatorics to a level that it can be managed
with very simple, fast executing algorithms.  A second key is
the ever increasing power of available computing.  It
is anticipated that the front end of Level~1 will
be implemented using Field Programmable Gate Arrays
(FPGA), while its back end will be a farm of 
Digital Signal Processors (DSP).  An option exists to use
a farm of general purpose processors for the back end.

Events which pass Level~1 will be sent to two more levels
of triggering, Levels~2 and~3.  Level~2 will use more refined
algorithms to perform essentially the same computation
as Level~1, while Level~3 will incorporate information
from additional detector subsystems.  
As for Level~1, the Level~2 and Level~3 algorithms will make
a decision based on evidence for a detached vertex, a strategy
which will give a high efficiency both for final states now
known to be important and for many final states
whose importance is not yet appreciated.
Both the Level~2 and~3 algorithms will be executed on a large farm 
of general purposes processors, of order 2500 computing nodes.  
These algorithms will be coded using the standard offline
software environment, with all of its available tools and
infrastructure.

  This picture places an important constraint on the
BTeV software: while the infrastructure must be powerful
enough to meet the complexity demands of offline work
it must also meet the speed demands of an online
system.  Both of these uses require a high level of
reliability and robustness.

%
%

\subsection{History of the BTEV Software}

Design studies for BTeV date back to the mid 1990's and
were originally performed using MCFast~\cite{MCFast},
a fast, parameterized detector simulation package.
After a few years, a Geant~3 based package, BTeVGeant, was
developed for detailed simulations, in particular for
those studies in which it is important to
simulate the production of new particles produced by the 
interaction of their parent particle with 
the detector and support materials.  

A suite of trigger, reconstruction and analysis
algorithms was developed to operate on the hits generated
by these packages.
Many of these algorithms are quite
sophisticated; for example the track fitting is done using
a Kalman filter and the reconstruction codes for the
Electromagnetic Calorimeter (ECal) are 
derived from the algorithms used by the Crystal Ball
and CLEO collaborations.  The mass and vertex
constrained fitting package is based on 
{\tt KWFIT}\cite{kwfit}. 
The most mature of these codes are the prototype codes for the Level~1 and
Level~2 trigger algorithms.

  These trigger, reconstruction and analysis codes were developed
very quickly and were available early in the design process.
This permitted BTeV to use high level physics metrics to evaluate
design changes.  For example, when the
amount of material in the pixel detector support structure
was increased, it was straight forward to evaluate, for
many final states, the degradation in vertex resolution, 
mass resolution and efficiency due to scattering, interactions 
and pair conversions in the additional material.  The
numbers presented in the BTeV proposal and related documents
were obtained using this software.

  With the exception of the prototype Level~1 trigger codes, none
of these codes were designed to be used in the long term.
They were designed to give detailed and precise but fast estimates
of the physics reach of detector variations.
To achieve these goals with the few people and short
time available, many assumptions and idealizations
were hard coded.  Most detector components, for example, are 
presumed to be perfectly aligned and no provision is made
to correct for misalignment; this turns many 3D problems into
1D or 2D problems, greatly simplifying the algebra and
coding times.
 The facilities for handling of
data and meta-data are primitive and will not scale to the
anticipated data volumes.  Finally, use of these codes
requires fluency with many magic words and phrases, documented
only by oral tradition.

    These codes, written in a mixture of Fortran, C and 
C++, have served BTeV well but a well planned and
well executed successor is needed take the next steps toward 
the physics goals of the experiment.

\subsection{Moving Toward the Future}

   BTeV is now working to design and implement a modern software 
infrastructure, written in C++,
in which the next generation of physics software will live.
It is expected that the first new physics codes will be written
starting about a year or two from now.  By that time we will
have defined the major components of the system and defined how
they interact with each other.  We will also have implemented
enough of the core infrastructure to allow quasi-independent
development of the physics codes and of the utilities
and services which they require.

The new software is not required to reuse any existing data
structures or code, although it may if that turns out to
be the best design decision.  While it is required to read files
written in the existing formats, it will, internally,
construct new style events before presenting the event data to
the users.

One can think of BTeV as part of the second generation 
of experiments using modern software, the first being 
those experiments which use predominantly use C++ and which are 
running now.  As part of the design process, BTeV is studying
the many successes and the few failures of the first generation
experiments.

\subsection{Understanding the User Community }
\label{sec:users}

One of the lessons learned from previous experiments is that
it is important to understand that the software will be used
by many people with a broad spectrum of skills.  By far
the largest part of the user community comprises
experienced physicists with little or no C++ experience
and little or no formal computer science training. In previous
generations of experiments, usually with FORTRAN based
software, such physicists wrote the vast
majority of the physics code and performed most of the
analyses. 
It is critical to the success of BTeV that these
physicists be able to get up to speed quickly and that they
not be marginalized by long learning curves.

  A second, and rapidly growing, class of users are those
experienced physicists with a small amount of C++ training
and practice.  One of the challenges of dealing with this
community is to convince them of the need for ongoing
education.  In particular they have a continuing needs to see
illustrations of good programming practice and to be
educated to avoid bad practices which have infiltrated
the communal bag of tricks.    Among these bad practices
are a reliance on variables with global scope,
inattention to const correctness, inappropriate use
of inheritance, unnecessary copies
of large objects hidden behind an opaque syntax, 
and the use of casts, particularly (void *), to defeat the type 
safety mechanisms, rather than designing to avoid the need for casts.
A particularly pernicious problem is to is a tendency to overuse 
newly learned skills and tricks, forcing them on problems for which
there are more natural solutions.

  We anticipate only a small handful of users who are both
experienced physicists and well trained, accomplished
C++ programmers.   These people need concise, well indexed
and well cross-referenced, documents which describe
the BTeV specific software.

  Another important class of users are new physicists,
inexperienced but energetic and enthusiastic, who have
the same broad spectrum of programming backgrounds as
do their experienced counterparts.  For all users,
but for these users in particular, there needs to be
cross-referencing from the code documentation to the 
corresponding physics documentation.

  One of the lessons reported by other experiments is that
most users start a new coding project by finding related
code and modifying it to suit the new task. 
(~In the \LaTeX\ source for this report, the author has included
some fragments which he copied from someone in the  mid-1980's!~)
Indeed, this is how most people learn both the experiment 
specific software tools and how to use a new computer language.  
Therefore early code fragments are extensively copied and any
poor design choices found in them will be widely propagated;
even outright errors will be widely propagated.  

   Therefore it is critical for the BTeV software team
to provide these early code fragments and to make people
aware that they exist. This is a natural mission for a
suite of tutorial examples.  In order to be widely accepted
the tutorials must implement solutions to real world 
problems which are encountered in the day to day life of
a working high energy physicist.
A corollary of this discussion is that the interfaces seen in 
these tutorials must be among the first designed.
A second corollary is that the tutorials must be available
very early and must be maintained throughout the development process.

\subsection{Outline of the Software Model}

  An early prototype of a new software suite for
BTeV is now available.
The design of this prototype has four components,
the framework, an event data model (EDM),
modules and services.
The physics codes will live in the modules, which
can respond to such occurrences as start of job, start of run, 
new event, end of run, end of job and others which are yet to be defined.
The job of the framework is to learn that some thing such
as start of a new run or a new event in memory has occured and
to then call the appropriate method of each module.  The order
in which the modules are called is specified by the user.
Modules may communicate with each other only via the EDM
and they may influence flow control by sending messages back
to the framework.  The so called services are present to
provide information and services to the modules and to the
framework.  Examples of services include a message logger, a geometry
manager, a calibration manager,  memory use monitors, event
timers and so on.  In the present design, input and output  
(IO) is done by a specialized set of modules.

  The existing prototype has a framework, a module
base class, some module concrete classes, a few prototype
services, a run time configuration facility based on the 
Run Control Parameter (RCP) system from the DZero experiment
and the rudiments
of an EDM, including data provenance.  The prototype also
has track, shower and vertex classes, collections of which 
live in the EDM.  The prototype IO module reads events 
created by BTeVGeant and reformats them into a new style
event.  It does not yet write events in the new style but
it can write out selected events in the old style.

   Two of the software engineers working part time on
BTeV spend the majority of their time supporting the CDF and
DZero Run~II software effort.  So the existing prototype code borrows
ideas, and some code fragments, the CDF and DZero Run~II software.
Their code was not reused outright because it contains compromises
needed to deal with their legacy software and legacy use patterns.

   Most of the effort to date has gone into understanding the 
interfaces among the major components.  Particular attention has
been paid to interfaces which will be seen frequently by
inexperienced C++ programmers, especially those interfaces 
which will be used heavily during data analysis.  Asking
a question as simple as ``how do I make a histogram of the
 momentum of all tracks'' touches on many interfaces.
For these interfaces 
the overriding design principle was to make the interface as simple
as possible.  Often this meant introducing complexity at a lower
level, an acceptable trade-off because experienced C++ 
designers are available to design and write the lower layers.
In some cases
the team could not produce a design with both the required
capabilities and a a truly simple interface.  In these cases
there are two design principles: the interfaces must conform to 
a small number of patterns and the use of these patterns must be easy to
teach to inexperienced users, even if they do not fully 
appreciate all of the details.
The hope is that, if the complex interfaces conform to a small number of 
patterns, new users can learn a difficult lesson once and apply
it many times.

  This plan, of concentrating first on the analysis interfaces
was adopted during 2002.  It arose from the observation,
discussed in section~\ref{sec:users}, that a small number of
early code fragments will be heavily copied and will set the
overall tone and quality of physics analysis software throughout
the experiment.  The design team
was encouraged to hear several reports at this CHEP in which
the speakers said that their design effort should have 
paid more attention to the analysis phase of the experiment.

  So far the prototype code is rigorous about type safety,
exception safety and const correctness.
The design team
is evaluating the SIUnits\cite{siunits} package and will soon
make a recommendation about its use.  The prototype code
is also being reviewed to ensure that it is thread safe; 
it is anticipated, based on the experience of CDF and DZero
in Run~II, that the code needed to access the calibration
database may well use threads.

The final piece in the existing prototype is a tutorial suite.

\section{THE TUTORIAL SUITE }

\subsection{The Mandate}

The mandate for the tutorial suite is to
allow all physicists, but particularly those with 
little or no C++ experience, to do useful work as quickly as possible.
And there are many related goals.
The tutorial suite must sell the new software
infrastructure to the collaboration, which means selling both
the choice of C++ and our design in particular.
It must teach good C++ practice, both
in general and in situations which are peculiar to the BTeV software.
It should
give an overview of all of the software tools supported by or
recommended by BTeV.  This overview should form an index into
the detailed documentation, including the documentation 
for the C++ language, for third party products, such as ROOT,
and for BTeV specific tools.

\subsection{How to Achieve these Goals}

To meet the mandate, tutorial examples must be chosen from real world
problems in the day to day life of a working high energy physicist.
Such problems range from occupancy maps of a subdetector,
to the inclusive momentum distribution of all tracks, 
to complete simulated analyses. When data is available the
examples should include real analyses.  Each of the examples must
always ``just work'' and each must produce
something concrete, such as a histogram or a formatted printout
which can be compared to a reference.  That reference must be
distributed along with with the example code.
Ideally the instructions should be no more complex than: 
check out the tutorial from CVS; gmake;
look at the histograms.

Each of the tutorial examples is accompanied by
narrative documentation because a reference manual alone would
be far from adequate.  Narrative documentation
is particularly important for the first few examples in which
many new ideas must be introduced. As much as possible the
narrative should start from familiar ideas and proceed comfortably
from there to the unfamiliar.  The narrative 
should spiral in toward the details of the problem, making a 
short first pass which gives an overview, adding 
details on successive passes.

  Consider for example, the existing first example.  This
loops over all of the reconstructed charged tracks in an
event to make
two histograms and one ntuple.  The overview immediately
answers four questions, 

\begin{enumerate}
\item Where do I specify the run time configuration information such
      as the number of events to read, the name of the input file
      and the name of the histogram file?
\item Where do I find the code which is called once at the start of the job,
      such as booking histograms?
\item Where do I find the code which is called once per event, such as
      filling histograms?
\item Where do I find the code which is called once at the end of the job?
\end{enumerate}

The narrative gives brief answers to these questions.  It then says
to look at the code which is called called once at the
start of the job and identify the lines which book the histograms.
The narrative then mentions that the histogram package is 
root and includes a link to the root documentation.   Next it says
to find the code which is called once per event
and find the code fragment, inside the track loop, which deals with the
properties of one track.   The narrative gives a few sentences
about what information is available about reconstructed
charged tracks, followed by a link to the detailed documentation
about these tracks.  The first pass concludes by pointing out
that the histograms are automatically written out at the end of
job.

Until this point, the narrative documentation has not mentioned
any of the words framework, module, EDM, C++, class, object, 
const\_iterator, template, handle, STL, vector and so on.
The subsequent passes of the narrative documentation briefly
introduce these ideas and include links to their
detailed documentation.  Usually this detailed documentation does 
not refer to a immediately to reference manual.  Instead it a refers 
to a narrative description which, in turn, has links to
the reference manuals.
 As each code fragment is discussed
in detail, the documentation mentions both the big ideas
from the point of view of the design of the BTeV software
and comments on any new C++ language or syntax elements which 
are encountered for the first time.  In the future these parts
of the narrative will also include links to an online C++ language
reference.

Writing the narrative documentation was an iterative process.
There were several reorganizations of the detailed documentation
so that it would be more natural to link it from the narrative
documentation of the tutorial.   There were also several reorganizations
after receiving feedback from BTeV physicists who had 
started to use the new software.

The second of the existing tutorial examples loops over all reconstructed
ECal showers in an event and makes some histograms of their
properties.  The narrative documentation for this example is
much shorter since it can refer back to the narrative documentation
for example~1 to discuss the big picture issues of frameworks
and modules and so on.  Whenever possible the narrative emphasizes the 
similarities  between looking at tracks and looking at showers.
As this documentation was written it required reorganization
of some of the material in the detailed documentation and some
of the narrative for the first example.

A few paragraphs earlier
it was pointed out that a reference manual alone is not sufficient.
But it is most certainly necessary.  The Milano group within
BTeV has developed software for control of the DAQ system
and monitoring of data quality 
in the next test-beam run for the BTeV pixel detector.  They
have used DoxyGen\cite{doxygen} to produce an online
reference manual for their system.  Based on this experience
the offline software team anticipates using DoxyGen, or a similar
product, to provide online reference manuals.  The narrative
documentation will then be updated with links into the reference
manuals.

At present each of the tutorial examples provides a set of
reference histograms and users are instructed to compare the
histograms from their test run against the reference.  
This step is intended to give the users confidence that they
have correctly compiled, linked and executed the software.
It has the unfortunate side effect that the least experienced
users are the first to discover 
many small bugs.  In the future each tutorial will
be compiled, linked and run as part of a nightly validation
suite.  At that time the output histograms will be compared
programatically to the reference and, if they are not identical, 
a message will be sent to the software czar.  In this
way it is hoped that new users will be better
shielded from undiscovered bugs.  At the presentation
of this talk at CHEP, several people from the audience encouraged
BTeV to do this as soon as possible as they found it invaluable
for their own experiments.

Part of the mandate is to sell C++ to the collaboration.
While most of the collaboration is accepting of C++,
and many are even enthusiastic, there remains
a small but vocal group of skeptics.
Their skepticism derives
from bad experiences, either their own or their colleagues',
with C++ based software on other experiments.
After talking with several of these people the design team concluded 
and that a well crafted set of tutorials would go a long
way toward alleviating most of their concerns.  

\subsection{A www Wish List}

An issue for which BTeV has a partly satisfactory 
solution is how to synchronize code 
documentation with  code versions. This is particularly 
important for the new software
which is changing rapidly.  At present all documentation, including
the web pages, are stored within the code repository in which the
code is stored.  This documentation is tagged with the same version
stamps as is the code and both are served to the web
from the main BTeV web site.  When one asks to see the documentation for
the tutorials, the reader is told about available code versions
and asked to pick one.  The documentation for that version will
then be shown.
In this way one, may change the documentation at the head of 
the repository to match code changes and not worry too much that it will
confuse someone who is still working with an earlier version
of the code. 

While this system can be used to ensure that documentation within
the tutorials is internally consistent, it does not stop someone who
is maintaining an unversioned web page, or a web page with
an independent version sequence, from linking to one of the
web pages inside of the tutorials.  As the code evolves there
is no reliable mechanism to ensure that the link is updated to
point to the new version of the documentation.
BTeV is interested in learning about more robust solutions to this problem.

In the narrative documentation
within the tutorials, short code fragments from the example
code are copied into the narrative documentation and 
are discussed in the following paragraphs.  At present these
code fragments are copied by hand from the source code file
into the .shtml file which holds the narrative documentation.
BTeV would like to have a tool which can extract appropriately
marked lines from the source file and drop it into the narrative
documentation.  This would help to keep the code and documentation
synchronized.  

\subsection{Looking Ahead}

  As BTeV continues to develop the new software, the tutorials
will be updated to match.  As the tutorials are updated,
and as people use them and give feedback, the design team will learn
which features people find easy to use and which they find hard,
which features are missing and needed immediately, which information
lives in the wrong place or is otherwise hard to find, 
and which features
are simply undocumentable or unteachable to new users.  These
lessons will be fed back into the design of the new code and into
the design of the tutorials.

  The long term goal is to have a complete offline software suite,
both the
infrastructure and the physics code,
available well before data taking starts in 2008.   Moreover
the major components need to be functioning well by 
2006 in order to support engineering runs on partially installed
subsystems.  The short term goals are more modest:  to prepare
the ground for the long term.  In the next six to 12 months
the design team expects to complete the design of the major
interfaces and to continue the work on the infrastructure code.
An important part of the work in this time will be to maintain
and extend the tutorial suite, so that it can be used to train
the physicists and computer scientists who will meet the long term goals.

\section{SUMMARY AND CONCLUSIONS}
 
  BTeV has begun the long process of designing and implementing
a modern offline software suite which will be used both in
the Level~2/3 trigger and for offline data processing.
In the initial stages the design team has focused on the 
interfaces which will be seen frequently by inexperienced 
C++ programmers, particularly during data analysis.  As soon
as prototype code was available the design team produced 
a suite of tutorials to teach the new software to other
the full spectrum of BTeV physicists.  These tutorials are 
an integral part of the BTeV software suite and they will
evolve along with the project.  The tutorials will serve
as a test bed for new ideas and will be integrated into
a nightly validation suite.  They will also serve as
an index to and overview of the detailed documentation 
for all of the collaboration's software.

\begin{acknowledgments}

The author would like to thank the conference organizers
for a well run and informative conference held in most
pleasant surroundings.  
This work was supported in part by Fermilab, which is operated
by the Universities Research Association, Inc. under Contract
No. De-AC02-76H03000 with the United States Department of Energy.

\end{acknowledgments}


\end{document}